# Analysis of Microprocessor Based Protective Relay's (MBPR) Differential Equation Algorithms

Bruno Osorno

**Abstract**— This paper analyses and explains from the systems point of view, microprocessor based protective relay (MBPR) systems with emphasis on differential equation algorithms. Presently, the application of protective relaying in power systems, using MBPR systems, based on the differential equation algorithm is valued more than the protection relaying based on any other type of algorithm, because of advantages in accuracy and implementation. MBPR differential equation approach can tolerate some errors caused by power system abnormality such as DC offset. *This paper shows that the algorithm is a system description based and it is immune from distortions such as DC-offset*. Differential equation algorithms implemented in MBPR are widely used in the protection of transmission and distribution lines, transformers, buses, motors, etc. The parameters from the system, utilized in these algorithms, are obtained from the power system current i(t) or voltage v(t), which are abnormal values under fault or distortion situations. So, an error study for the algorithm is considered necessary.

—————————— ◆ ——————————

## 1 INTRODUCTION

Microprocessor based protective relays are developed on the basis of early computer relaying devices. They, in turn, inherit some of the computer relays' functions in both hardware and software. Although they are now physically different, the fundamental principles and algorithms on which they operate remain similar.

The use of microprocessors in protection relays started a new era in which the MBPR dominates today's world of protection relays [1], [2], [3]. It overcomes the three main problems: _speed, accuracy and cost_, all of which hinder the extensive application of computer relaying devices. With the advent of more and more powerful microprocessors and the development of analytical algorithms, microprocessor based protective relays may provide a considerable upgrade in protection relaying speed, accuracy, reliability, and cost effective schemes to the modern sophisticated power systems.

## 2 TYPICAL HARDWARE ARCHITECTURE

With the development of several decades, a typical pattern of MBPRs can be found in most relay centers as shown in Fig. 1, [3].

Figure 1 is the basic block diagram of a MBPR showing the overall functional arrangement of hardware within the relay. Line voltage and current from PTs and CTs, are connected to isolating transformers, anti-aliasing filters and sample circuits. AC inputs are connected through a multiplexer to the analog conversion subsystem. An A/D converter places instantaneous samples of these ac signals

• *B. Osorno is with the Electrical and Computer Engineering Department, California State University, Northridge, CA 91330.*

in the microprocessor memory. Status and contact inputs are also scanned. However, what a MBPR looks like in terms of hardware depends on the functions needed for a specific protection relaying scheme. Some blocks may not exist in the figure, and others may become more or less significant. The configuration is applicable to most MBPR systems available at this time [1], [3].

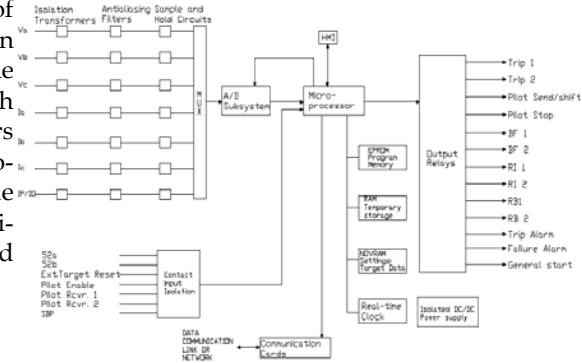

Fig. 1 Simplified Diagram of a Typical Microprocessor Based Relay

The exception is the class of relays that use the principle of traveling waves. A digital relay is required to receive one or more power systems operational parameters. For example, an over-current relay may be required to use data of three phase currents and the residual current in a circuit. On the other hand, a bus differential relays might require up to 30 inputs. Voltage and current are the operational parameters that are mostly used in protection relays. These parameters ranging up to hundreds of kilovolts and thousands of amps need to be transmitted into standard secondary levels of CT or PT.

The lower level signals from the transducers are applied to the analog input system The purpose of this sys-tem is: (a) to isolate the relay from the power system, providing protection from transient overvoltage, (b) to attenuate high frequencies sufficiently minimizing the consequences of aliasing, (c) to reduce the level of voltages and current to equivalent voltage signals.

In Fig. 1 the outputs of the analog input subsystem are applied to the analog interface subsystem. This subsystem includes sample and hold units, converters and multiplexers. The functions of sampling, converting and multiplexing can be performed using the designer's own philosophy. For example, analog signals can be sampled and held as voltages across capacitors. A multiplexer can then apply each voltage to digital values and input it to the microprocessor. Alternatively each signal can be sampled and converted to digital values using dedicated A/D converters. A multiplexer then reads information sequentially into the microprocessor.

The status of circuit breakers and isolators, relay targets and voltage sense information from the power system is provided to the relay via the digital input systems. An auxiliary power source and sensing mechanism is used to sense the status of contacts. Since transient voltage can be experienced on the digital input wiring, isolating arrangements, shielding up the input wiring and transient protections are provided.

The output of a digital relay is transmitted to the power system through the digital output subsystem. A relay may be required to provide up to ten digital outputs. Suitable isolation of the microprocessor output circuits from the power system and protection from transient must be included.

It is important to save the raw data, record it for post fault analysis and use it in the future. RAM (random access memory) is used for temporary storage of transient data and it is freed for the next occurrence of a transient.

The relay program logic is stored in ERAM (Erasable RAM), the CPU, and the registers as a unit to execute the program one statement at a time. *A watching dog chip resets the system in case of a software malfunction.* Analog data is sampled continuously to measure instantaneous values of the signal from the power system. To obtain constant results, the aperture must be small. At present sampling rates between 240 Hz and 2 kHz are being used. The most common form of sampling consists of taking a sample every ∆T second. In this case, the sampling frequency is specified as $f_s = \frac{1}{\Delta T}$Hz. For example for a 2 kHz frequency ($f_s$), ∆T could be 0.5 ms.

## 3 DIFFERENTIAL EQUATION ALGORITHM

Differential equation algorithms are based on the model of a power system rather than on the model of a signal. The method using a simple single phase model of a faulted line is discussed.

### 3.1 R-L model Shorter Window Algorithm

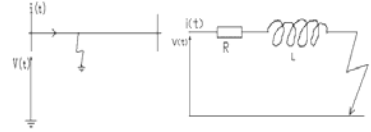

Fig. 2 Representation of Transmission Line Using Lumped-Series Circuit Parameters

As we know, voltage in a line can be expressed by the following equation [3];

$$V(t) = R \cdot i(t) + L \frac{di}{dt} \quad \ldots \quad (1)$$

where: R- line impedance.
L- line reactance.

Since we have only two components, R and L, which are used to describe the line system, need to be determine in the equation, and they are totally dependent on the line features. So, a line can be modeled as a R-L network.

By integrating over two successive time periods, we have

$$\int_{t0}^{t1} v(t)dt = R \int_{t0}^{t1} i(t)dt + L[i(t_1) - i(t_0)] \ldots \quad (2)$$

$$\int_{t1}^{t2} v(t)dt = R \int_{t1}^{t2} i(t)dt + L[i(t_2) - i(t_1)] \ldots \quad (3)$$

The integrals in (2) and (3) must be approximated from the sampled values. If the samples are equally spaced at an interval $\Delta t$ and the trapezoidal rule is used for the integrals, Viz.

$$\int_{t_0}^{t_1} V(t)dt = \frac{\Delta t}{2}[V(t_1) + V(t_0)] = \frac{\Delta t}{2}[V_1 + V_0] \ldots \quad (4)$$

Then (2) and (3) can be written for samples at $k, k+1 \text{ and } k+2$ as

$$\begin{bmatrix} \frac{\Delta t}{2}(i_{k+1} + i_k) & (i_{k+1} - i_k) \\ \frac{\Delta t}{2}(i_{k+2} + i_{k+1}) & (i_{k+2} - i_{k+1}) \end{bmatrix} \begin{bmatrix} R \\ L \end{bmatrix}$$

$$= \begin{bmatrix} \frac{\Delta t}{2}(v_{k+1} + v_k) \\ \frac{\Delta t}{2}(v_{k+2} + v_{k+1}) \end{bmatrix} \ldots \quad (5)$$

If the integrals are evaluated using the trapezoidal rule then a number of different algorithms may be given by varing the number of samples in the two intervals. The minimum is 3 samples total resulting in

$$L = \frac{\Delta t}{2}\left[\frac{(i_k+i_{k-1})(V_{k-1}+V_{k-2}) - (i_{k-1}+i_{k-2})(V_k+V_{k-1})}{(i_k+i_{k-1})(i_{k-1}-i_{k-2}) - (i_k-i_{k-1})(i_{k-1}+i_{k-2})}\right] \ldots \quad (6)$$

$$R = \left[\frac{(V_k+V_{k-1})(i_{k-1}+i_{k-2}) - (V_{k-1}+V_{k-2})(i_k+i_{k-1})}{(i_k+i_{k-1})(i_{k-1}-i_{k-2}) - (i_{k-1}+i_{k-2})(i_k-i_{k-1})}\right] \ldots \quad (7)$$

Recall that a complex division is required to calculate the impedance using the estimated phasor in the waveform approach. An obvious advantage of the differential equation algorithm is that the DC-offset is not an error signal since it satisfies the differential equation and hence does not have to be removed.

### 3.2 Proper selection of intervals for rejecting particular harmonic components

The algorithm described by (6) and (7) is a short window algorithm and is not as selective as a longer window algorithm [4]. A longer window of data will sharpen the frequency response of the algorithm and more accuracy of estimates of interests (R and L) will be obtained. To extend the differential equation algorithm to a longer window, we can make the intervals [$t_0,t_1$], [$t_1,t_2$] longer and properly choose the length of the intervals so that certain harmonics are rejected. This is a type of form in terms of "*longer window*". If the intervals contain a number of samples, the trapezoidal approximation to each of the intergrals will contain sums of a number of samples.

Assume that a current waveform i(t) is periodic with the period of T, its Fourier expression is

$$i(t) = \frac{a_0}{2} + a_1 \cos\omega_0 t + a_2 \cos 2\omega_0 t + a_3 \cos 3\omega_0 t + \ldots$$
$$+ b_1 \sin\omega_0 t + b_2 \sin 2\omega_0 t + b_3 \sin 3\omega_0 t + \cdots \quad (8)$$

If the highest harmonic contained in the waveform i(t) is N, the last equation would be reduced to

$$i(t) = c_0 + \sum_{m=1}^{N} C_m \cos(m\omega_0 t + \theta_m) \ldots \quad (9)$$

Where:
$c_0 = \frac{a_0}{2}$,
$C_m = \sqrt{a_m^2 + b_m^2}$ and $\theta_m = \tan^{-1}\frac{b_m}{a_m}$

Now we integrate equation 9 from the selected range $t_1=0$ to $t_2=\alpha/\omega_0$, and do the integration:

$$I_1 = \int_{t_1=0}^{t_2=\alpha/\omega_0} i(t) = \int_0^{\alpha/\omega_0} C_0 dt + \sum_{m=1}^{N} \int_0^{\alpha/\omega_0} C_m \cos(m\omega_0 t + \theta_m) dt$$

As indicated in the above equation, if $I_{n1}$ is the integration of the nth harmonic over the period from $t_1=0$ to $t_2=a/\omega_0$, then

$$I_{n1} = \int_0^{\alpha/\omega_0} C_n \cos(n\omega_0 t + \theta_n) dt$$
$$= \frac{C_n}{n\omega_0} \sin[n\alpha + \theta_n - \sin\theta_n] \ldots \quad (10)$$

Similarly, if we integrate i(t) from $t_3=\pi/n\omega_0$ to $t_4=(\pi/n+\alpha)/\omega_0$, we have

$$I_{n2}$$
$$= \int_{\pi/n\omega_0}^{(\pi/n+\alpha)/\omega_0} C_n \cos(n\omega_0 t + \theta_n) dt$$
$$= \frac{C_n}{n\omega_0}[-\sin(n\alpha + \theta_n) + \sin\theta_n] \ldots \quad (11)$$

It will be evident from equation 10 and 11 that $I_{n1}+I_{n2}=0$. This is of course true for all harmonics (n). In essence, this means that any nth harmonic and its multiples can be filtered out from the waveform i(t), by simply adding the intergrals taken once over the limits $t_1=0$ to $t_2=\alpha/\omega_0$ and $t_3=\pi/n\omega_0$ to $t_4=(\pi/n+\alpha)/\omega_0$. By following the above process, i(t) becomes a pure fundamental frequency signal, on which we can gain a quality of estimates of R and L of the line.

### 3.3 Counting algorithm for implementation of longer window

Equations (6) and (7) are applied with the short window algorithm [1], [3], which is not as selective as a long window algorithm [1], [3]. The simple way to extend (6) and (7) to a long window application is to make the intervals [$t_0,t_1$],[$t_1,t_2$] longer. Also simultaneously, properly selecting the intervals, certain harmonics can be rejected. One of the approachs to be introduced for a longer window is to do the trapezoidal integration over the interval between adjacent samples and then obtain an over-defined set of equations in the form.

$$\begin{bmatrix} \frac{\Delta t}{2}(i_{k+1} + i_k) & (i_{k+1} - i_k) \\ \frac{\Delta t}{2}(i_{k+2} + i_{k+1}) & (i_{k+2} - i_{k+1}) \\ . & . \\ . & . \\ . & . \\ \frac{\Delta t}{2}(i_{k+N} + i_{k+N-1}) & (i_{k+N} - i_{k+N-1}) \end{bmatrix} \begin{bmatrix} R \\ L \end{bmatrix}$$
$$= \begin{bmatrix} \frac{\Delta t}{2}(v_{k+1} + v_k) \\ \frac{\Delta t}{2}(v_{k+2} + v_{k+1}) \\ . \\ . \\ . \\ \frac{\Delta t}{2}(v_{k+N} + v_{k+N-1}) \end{bmatrix} \ldots \quad (12)$$

The least square solution of the over-defined equations would involve a large number of multiplications. To avoid some of these problems, a technique of using a sequence of estimates, each of which is obtained from the three-sample algorithm, has been developed. Suppose a region of the R-L plane is chosen to correspond to zero-one protection of the line. If the result of the three-point calculation 6 and 7 lies in the characteristic, a counter is indexed by one. If the computed value lies outside the characteristic, the counter is reduced by one. With a threshold of four on the counter, i.e. a trip signal cannot be issued unless the counter is at four, a minimum of six consecutive samples are required. The window length is thus increased by increasing the counter threshold.

Although ingenious, the counting schemes are difficult to compare with other long window algorithms in terms of frequency response. An additional problem in giving the frequency response of the differential-equation algorithms is that there are two signals involved in the creation of frequency response. To obtain some comparisons, Fig. 3 shows a frequency response obtained for the average of consecutive three-sample results of (6) and (7) spanning half-cycle and full-cycle windows at a sampling rate of twelve samples per cycle. In computing Figure 3 it was assumed that the current was the true fundamental frequency current but that the voltage signal varied in frequency. In essence, Figure 3 is the frequency response if the averaged numerator of equations (6) and (7) with the denominator help at the fundamental frequency values. The frequency response in Fig. 3. can roughly be compared with the earlier responses.

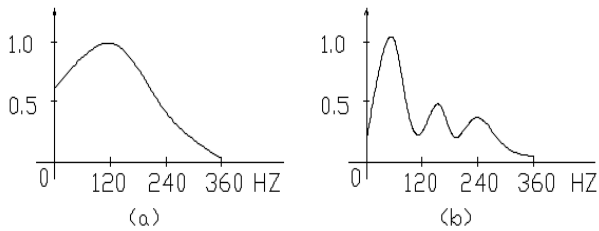

Fig. 3 Frequency Response of The Average of Consecutive Three-Sample Calculations. (a) One-half cycle window at twelve samples per cycle. (b) Full-cycle window at twelve samples per cycle

### 3.4 Error analysis

The property identified in describing the frequency response of the differential-equation algorithms, i.e. that the correct R and L are obtained as long as $v(t)$ and $i(t)$ satisfy the differential equation, is a major strength of these algorithms [1]. The exponential DC-offset in the current satisfies the equation if the correct values of R and L are used, so that it does not have to be removed. The voltage seen by the relay just after inception has non-fundamental fequency components caused by the power system itself. If the faulted line can accurately be modeled as a series R-L line, then the current will respond to these non-fundamental frequency components in the voltage and no errors in estimating R and L will result. As long as the R-L representation is correct then the only source of errors is in the measurement of $v(t)$ and $i(t)$. To examine the effect of such errors in both voltage and current, let the measured current and voltage be denoted by $i_m(t)$ and $v_m(t)$ respectively, where

$$i_m(t) = i(t) + \varepsilon_i(t) \quad \text{and} \quad v_m(t) = v(t) + \varepsilon_v(t)$$

and $i(t)$ and $v(t)$ satisfy the differential equation. The current $i_m(t)$ satisfies the differential equation

$$Ri_m(t) + L\frac{di_m(t)}{dt} = v(t) + R\varepsilon_i(t) + L\frac{d\varepsilon_i(t)}{dt} \ldots \quad (13)$$

and the current $i_m(t)$ and voltage $v_m(t)$ are related by:

$$i_m(t) + L\frac{di_m(t)}{dt} = v_m(t) + R\varepsilon_i(t) + L\frac{d\varepsilon_i(t)}{dt} - \varepsilon_v(t) \ldots \quad (14)$$

The measured voltage and current then satisfy a differential equation with an error term $\varepsilon_i(t)$ or $\varepsilon_v(t)$ which is made up of the voltage error plus a processed current error term. The latter is similar to the error contained at the output of a mimic circuit with the current error as an input. Fig. 3. then can finally be interpreted as the response of the algorithm to the entire error term in equation 14 assuming that the sum of the last three terms on the right hand side of (14) is thought of as a signal $cos(\omega t)$. The last term in (14) also makes it clear that error signals that satisfy the differential equation do not contribute a net error to equation (14).

There is an additional subtlety in the three-sample differential-equation algorithm. The denominator of (6) and (7) is not a constant but rather a function of time which has maxima and minima. The denominator can be simplified to

$$(i_{k+1} + i_k)(i_{k+2} - i_{k+1}) - (i_{k+2} + i_{k+1})(i_{k+1} - i_k) = 2(i_{k+1}^2 - i_k i_{k+2})$$

If we assume that

$$i_{k+1} = I\cos(\omega_0 t) - I\cos(\omega_0 t_0)e^{-\frac{R}{L}(t-t_0)}$$

and use twelve samples per cycle ($\theta = 30^0$) with a line time constant of 40 ms, then

$$2(i_{k+1}^2 - i_k i_{k+2}) = I^2 \left[0.5 - 0.5384\cos(\omega_0 t + 7.41^0)\cos(\omega_0 t_0)e^{-\frac{R}{L}(t-t_0)}\right]$$

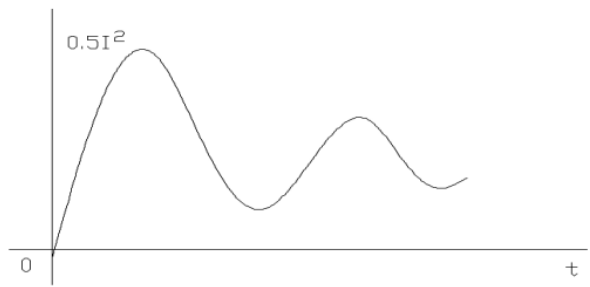

Fig. 4. The Denominator of 6 and 7 for Maximum Offset In the Current.

The denominator is shown in Fig. 4. as a function of the time of the $k^{th}$ sample for the case of maximum offset ($\omega_0 t_0 = 0$). When the denominator is small, the error terms from (14) are amplified. In the limit as the denominator becomes zero the estimates are unacceptably sensitive to even the smallest error terms. A counting algorithm will deal with such poor estimates by indexing down because the estimate is not in the characteristic. The net effect is then only a delay in issuing the trip signal. Since the denominator is a constant if the offset is absent, it can be seen that the supposed immunity of the diferential-equation algorithm to offset is a bit of a myth.

## 4 CONCLUSIONS

In the calculations of R & L in the differential algorithm, the corrupted waveforms would add error to the results of solving differential equations. *So in the analysis of the algorithm, it is important to carry out an error study.*

The realization of the differential equation algorithm needs burdensome calculations, which lead to slow reaction to the systems' fault.

With the availability of increasingly powerful microprocessors, the relaying with the differential equation algorithms will become stronger and the cost would be lower.

In recent years, a new algorithm is being developed: Traveling Wave Algorithm. Although it has not found its use in practice, it has displayed an extensive perspective in the applications of MBPR, especially in the protection relaying of long-distance transmission lines. In terms of hardware, nearly all of the MBPR share the hardware of Fig.1 including protective relays using differential algorithms.

After discussing the differential equation algorithm, we have shown that it is a system description based algorithm and it is immune from distortions such as DC-offset. Due to such feature, it can produce more accurate parameters (R & L) of a system. So, it is widely used in various protective relaying schemes of power systems.

## REFERENCES

[1] IEEE Tutorial Course, "Microprocessor Relays and Protection Systems". *The Institute of Electrical and Electronics Engineers, INC.*, 1987

[2] IEEE Tutorial Course, "Computer Relaying", *The Institute of Electrical and Electronics Engineers, INC.*, 1979.

[3] A. G.Phadke, J. S.Thorp, "Computer Relaying for Power Systems", *Research Studies Press LTD.*,1988

[4] A.T.Johns and S.K.Salman, Protection for Power Systems, Peter Peregrinus Ltd., 1977

[5] V. Gurevich, *Electric Relays Principles and Applications*, Taylor & Francis Group, LLC, 2006.

[6] W. A. Elmore, *Protective Relaying Theory and Application*, Second Edition, Marcel Dekker, INC., 2003

[7] S. H. Horowitz, *Protective Relaying for Power Systems*, IEEE PRESS, the Institute of Electrical and Electronics Engineers, INC., 1980.

[8] S. H. Horowitz, *Protective Relaying for Power Systems II*, IEEE PRESS, the Institute of Electrical and Electronics Engineers, INC., 1992.

[9] A.G.Phadke, M.Ibrahim, T.Hlibka, "Fundamental Basis for Distance Relaying with Symmetrical components" IEEE Transactions on Powers Apparatus and Systems, Issue 2, pp635-646, 1977

[10] J.G.Gilbert and R.J.Shovlin, "High Speed Transmission Line Fault Impedance Calculation Using Dedicated Microcomputer" *IEEE Transactions on Power Apparatus and Systems*, Vol. PAS-94, No 3, May/June 1975.

[11] D. Baigent and E. Lebenhaft, "Microprocessor-Based Protection Relays: Design and Application examples" IEEE transaction on Industry Applications, Vol. 29, No I, January/February 1993.

**Bruno Osorno** is an associate professor of Electrical and Computer Engineering at California State University Northridge. He has been the lead faculty member of the power systems option and is the director of the protection, power electronics and electrical machines laboratories. He has published over 25 papers related to power systems and education. His research interests are modeling of solar photovoltaic systems, modeling of power electronics, and recently design, implementation and education for the smart grid and microprocessor based relays used in protection.